\input harvmac.tex
\input epsf.tex

\def\figin{\epsfcheck\figin}\def\figins{\epsfcheck\figins}
\def\epsfcheck{\ifx\epsfbox\UnDeFiNeD
\message{(NO epsf.tex, FIGURES WILL BE IGNORED)}
\gdef\figin##1{\vskip2in}\gdef\figins##1{\hskip.5in}
\else\message{(FIGURES WILL BE INCLUDED)}%
\gdef\figin##1{##1}\gdef\figins##1{##1}\fi}
\def\DefWarn#1{}
\def\figinsert{\goodbreak\midinsert}
\def\ifig#1#2#3{\DefWarn#1\xdef#1{fig.~\the\figno}
\writedef{#1\leftbracket fig.\noexpand~\the\figno}%
\figinsert\figin{\centerline{#3}}\medskip\centerline{\vbox{\baselineskip12pt
\advance\hsize by -1truein\noindent\footnotefont{\bf
Fig.~\the\figno:} #2}}
\bigskip\endinsert\global\advance\figno by1}

\def\cz{{\cal Z}}


\lref\HasenfratzAB{
  P.~Hasenfratz and F.~Niedermayer,
  Phys.\ Lett.\  B {\bf 245}, 529 (1990). See also
  P.~Hasenfratz, M.~Maggiore and F.~Niedermayer,
  Phys.\ Lett.\  B {\bf 245}, 522 (1990).
}

\lref\afs{
  G.~Arutyunov, S.~Frolov and M.~Staudacher,
  JHEP {\bf 0410}, 016 (2004)
  [arXiv:hep-th/0406256].
}
\lref\bds{
  Z.~Bern, L.~J.~Dixon and V.~A.~Smirnov,
  Phys.\ Rev.\  D {\bf 72}, 085001 (2005)
  [arXiv:hep-th/0505205].
}

\lref\FrolovQE{
  S.~Frolov, A.~Tirziu and A.~A.~Tseytlin,
  Nucl.\ Phys.\  B {\bf 766}, 232 (2007)
  [arXiv:hep-th/0611269].
}
\lref\kruczwil{
 M.~Kruczenski,
  JHEP {\bf 0212}, 024 (2002)
  [arXiv:hep-th/0210115].
  }

\lref\bes{
  N.~Beisert, B.~Eden and M.~Staudacher,
  J.\ Stat.\ Mech.\  {\bf 0701}, P021 (2007)
  [arXiv:hep-th/0610251].
}
\lref\bhl{
  N.~Beisert, R.~Hernandez and E.~Lopez,
  JHEP {\bf 0611}, 070 (2006)
  [arXiv:hep-th/0609044].
}
\lref\es{
  B.~Eden and M.~Staudacher,
  J.\ Stat.\ Mech.\  {\bf 0611}, P014 (2006)
  [arXiv:hep-th/0603157].
}

\lref\jpms{ J.~Polchinski and M.~J.~Strassler,
  JHEP {\bf 0305}, 012 (2003)
  [arXiv:hep-th/0209211].
}

\lref\cusppolyakov{
  A.~M.~Polyakov,
  Nucl.\ Phys.\  B {\bf 164}, 171 (1980).
}
\lref\jmwilson{
  J.~M.~Maldacena,
  Phys.\ Rev.\ Lett.\  {\bf 80}, 4859 (1998)
  [arXiv:hep-th/9803002].
}

\lref\gkp{
  S.~S.~Gubser, I.~R.~Klebanov and A.~M.~Polyakov,
  Nucl.\ Phys.\  B {\bf 636}, 99 (2002)
  [arXiv:hep-th/0204051].
}

\lref\BennaND{
  M.~K.~Benna, S.~Benvenuti, I.~R.~Klebanov and A.~Scardicchio,
  Phys.\ Rev.\ Lett.\  {\bf 98}, 131603 (2007)
  [arXiv:hep-th/0611135].
   A.~V.~Kotikov and L.~N.~Lipatov,
  Nucl.\ Phys.\  B {\bf 769}, 217 (2007)
  [arXiv:hep-th/0611204].
  L.~F.~Alday, G.~Arutyunov, M.~K.~Benna, B.~Eden and I.~R.~Klebanov,
  JHEP {\bf 0704}, 082 (2007)
  [arXiv:hep-th/0702028].
  I.~Kostov, D.~Serban and D.~Volin,
  arXiv:hep-th/0703031.
}

\lref\factorization{
  J.~C.~Collins, D.~E.~Soper and G.~Sterman,
  Adv.\ Ser.\ Direct.\ High Energy Phys.\  {\bf 5}, 1 (1988)
  [arXiv:hep-ph/0409313].
}

\lref\costathree{
  L.~Cornalba, M.~S.~Costa and J.~Penedones,
  arXiv:0707.0120 [hep-th].
  }

  \lref\costa{
  L.~Cornalba, M.~S.~Costa, J.~Penedones and R.~Schiappa,
  Nucl.\ Phys.\  B {\bf 767}, 327 (2007)
  [arXiv:hep-th/0611123],
  arXiv:hep-th/0611122.
  }

\lref\freedman{
  E.~D'Hoker, D.~Z.~Freedman, S.~D.~Mathur, A.~Matusis and L.~Rastelli,
  Nucl.\ Phys.\  B {\bf 562}, 353 (1999)
  [arXiv:hep-th/9903196].
}

\lref\BelitskyEN{
  A.~V.~Belitsky, A.~S.~Gorsky and G.~P.~Korchemsky,
  Nucl.\ Phys.\  B {\bf 748}, 24 (2006)
  [arXiv:hep-th/0601112].
}

\lref\manypartons{
  A.~V.~Belitsky, A.~S.~Gorsky and G.~P.~Korchemsky,
  Nucl.\ Phys.\  B {\bf 667}, 3 (2003)
  [arXiv:hep-th/0304028].
}
\lref\kruczmany{
  M.~Kruczenski,
  JHEP {\bf 0508}, 014 (2005)
  [arXiv:hep-th/0410226].
}


\lref\GrossCS{
  D.~J.~Gross and F.~Wilczek,
  Phys.\ Rev.\  D {\bf 9}, 980 (1974).
 H.~Georgi and H.~D.~Politzer,
  Phys.\ Rev.\  D {\bf 9}, 416 (1974).
}

\lref\Korchsudakov{ G.~P.~Korchemsky and A.~V.~Radyushkin,
  Phys.\ Lett.\  B {\bf 171}, 459 (1986);
G. P. Korchemsky, and A. V. Radyushkin, Nucl. Phys. {\bf B283},
342 (1987);
  G.~P.~Korchemsky,
  Phys.\ Lett.\  B {\bf 220}, 629 (1989).
  S.~V.~Ivanov, G.~P.~Korchemsky and A.~V.~Radyushkin,
  Yad.\ Fiz.\  {\bf 44}, 230 (1986)
  [Sov.\ J.\ Nucl.\ Phys.\  {\bf 44}, 145 (1986)].
}

\lref\cuspcascading{
  M.~Kruczenski,
  Phys.\ Rev.\  D {\bf 69}, 106002 (2004)
  [arXiv:hep-th/0310030].
}
\lref\KorchemskySI{
  G.~P.~Korchemsky,
  Mod.\ Phys.\ Lett.\  A {\bf 4}, 1257 (1989).
  G.~P.~Korchemsky and G.~Marchesini,
  Nucl.\ Phys.\  B {\bf 406}, 225 (1993)
  [arXiv:hep-ph/9210281].
}

\lref\dolanweakcalc{
  F.~A.~Dolan and H.~Osborn,
  Nucl.\ Phys.\  B {\bf 629}, 3 (2002)
  [arXiv:hep-th/0112251].
}

\lref\sudakovreferences{ V. Sudakov, Sov. Phys. JETP {\bf 3} , 65
(1956). R. Jackiw, Ann. Phys. (N.Y.) {\bf 48}, 292 (1968).
   A.~H.~Mueller,
  Phys.\ Rev.\  D {\bf 20}, 2037 (1979).
   J.~C.~Collins,
  Phys.\ Rev.\  D {\bf 22}, 1478 (1980).
 A.~Sen,
  Phys.\ Rev.\  D {\bf 24}, 3281 (1981).
  }
  \lref\sudakovreview{
  J.~C.~Collins,
  ``Sudakov form factors,''
  Adv.\ Ser.\ Direct.\ High Energy Phys.\  {\bf 5}, 573 (1989)
  [arXiv:hep-ph/0312336].
}

\lref\msterman{
 L.~Magnea and G.~Sterman,
  Phys.\ Rev.\  D {\bf 42}, 4222 (1990).
 G.~Sterman and M.~E.~Tejeda-Yeomans,
  Phys.\ Lett.\  B {\bf 552}, 48 (2003)
  [arXiv:hep-ph/0210130].
  }

\lref\catani{
  S.~Catani,
  Phys.\ Lett.\  B {\bf 427}, 161 (1998)
  [arXiv:hep-ph/9802439].
}

\lref\sceft{ C.~W.~Bauer, S.~Fleming and M.~E.~Luke,
  Phys.\ Rev.\  D {\bf 63}, 014006 (2001)
  [arXiv:hep-ph/0005275].
  C.~W.~Bauer, S.~Fleming, D.~Pirjol and I.~W.~Stewart,
  Phys.\ Rev.\  D {\bf 63}, 114020 (2001)
  [arXiv:hep-ph/0011336].
  C.~W.~Bauer, D.~Pirjol and I.~W.~Stewart,
  Phys.\ Rev.\ Lett.\  {\bf 87}, 201806 (2001)
  [arXiv:hep-ph/0107002].
 C.~W.~Bauer, D.~Pirjol and I.~W.~Stewart,
  Phys.\ Rev.\  D {\bf 65}, 054022 (2002)
  [arXiv:hep-ph/0109045].
  A.~V.~Manohar,
  Phys.\ Rev.\  D {\bf 68}, 114019 (2003)
  [arXiv:hep-ph/0309176].
}


\lref\BassoNK{
  B.~Basso and G.~P.~Korchemsky,
  Nucl.\ Phys.\  B {\bf 775}, 1 (2007)
  [arXiv:hep-th/0612247].
}

\lref\kris{
  P.~Y.~Casteill and C.~Kristjansen,
  arXiv:0705.0890 [hep-th].
}

\lref\zamolzamol{  A.~B.~Zamolodchikov and A.~B.~Zamolodchikov,
  Annals Phys.\  {\bf 120}, 253 (1979).
  A.~B.~Zamolodchikov and A.~B.~Zamolodchikov,
  Nucl.\ Phys.\  B {\bf 133}, 525 (1978)
  [JETP Lett.\  {\bf 26}, 457 (1977)].
}

\lref\ultimokr{
  M.~Kruczenski, R.~Roiban, A.~Tirziu and A.~A.~Tseytlin,
  arXiv:0707.4254 [hep-th].
}

\lref\aldaymaldacena{
  L.~F.~Alday and J.~Maldacena,
  JHEP {\bf 0706}, 064 (2007)
  [arXiv:0705.0303 [hep-th]].
}


\Title{\vbox{\baselineskip12pt
}}
{\vbox{\centerline{ Comments  on operators with large spin }}}
\bigskip
\centerline{ Luis F. Alday$^{a,b}$ and Juan Maldacena$^b$}
\bigskip
\centerline{\it $^a$Institute for Theoretical Physics and Spinoza
Institute} \centerline{Utrecht University, 3508 TD Utrecht, The
Netherlands}

\centerline{ \it  $^b$School of Natural Sciences, Institute for
Advanced Study} \centerline{\it Princeton, NJ 08540, USA}

\vskip .3in \noindent
 We consider high spin operators. We give a general argument for
 the logarithmic scaling of their anomalous dimensions which is based on the
 symmetries of the problem. By an analytic continuation we can also see the origin of the
 double logarithmic divergence in the Sudakov factor. We show that the cusp anomalous dimension is
 the energy density for a flux configuration of the gauge theory on $AdS_3 \times S^1$.
We then focus on operators in ${\cal N}=4$ super Yang Mills which
carry large spin and SO(6) charge and show that in a particular
limit their properties are described in terms of a bosonic O(6)
sigma model. This can be used to make certain all loop
computations in the string theory.


 \Date{ }


\newsec{Introduction}

In this paper we focus on
  two issues. First we discuss how the so called ``cusp anomalous dimension'',
   $f(\lambda)$, appears in
various computations.  Namely in the dimension of high spin operators
and in lightlike Wilson loops with a cusp.
 These are well known relations
 and our only objective here is to give a
  different perspective on these issues.  First we
   give a general argument for the logarithmic behavior of
the anomalous dimension of high spin operators $\Delta - S =
f(\lambda) \log S $ \refs{\GrossCS,\KorchemskySI} which is based on the symmetries
of the problem. Then we
argue that the Sudakov form factor for two light-like
particles has a behavior $ e^{ - { f(\lambda) \over 4 }  ( \log
\mu)^2 }$  as a function of the IR cutoff
\refs{\sudakovreferences,\Korchsudakov,\msterman,\catani}
 (for a review see \sudakovreview ).
 This factor gives the leading IR
behavior  when we consider the exclusive scattering of colored
particles and it is an important ingredient in the computation of
amplitudes \factorization . Both of these properties follow from
symmetries of the theory plus the fact that we have gauge
fluxes.

For the particular case of planar ${\cal N}=4$ super Yang Mills,
 exact results were derived using integrability. In particular,
an exact integral equation was written whose solution gives the
cusp anomalous dimension for arbitrary $\lambda$ \bes . This
equation was analyzed further in \BennaND .

In the second part of this paper we consider the question of high
spin operators in ${\cal N}=4$ super Yang Mills that carry spin $S$
and one of the SO(6) charges, $J$, in the large $S,J$ limit such
that
 $ J/(\log S)$ is kept finite. In that case one can show that the
 anomalous dimension continues to have a logarithmic scaling
 \eqn\anomdim{
  \Delta - S = \left[ f(\lambda) + \epsilon (
\lambda,  { J \over \log S } ) \right] \log S } This type of
operators were studied in \FrolovQE\ (see also \BelitskyEN ),
where the function $\epsilon$ was computed up to one loop in the
strong coupling expansion. We derive an exact expression for
$\epsilon$ in a suitable limit. We do this by noticing that the
full $AdS_5 \times S^5$ sigma model reduces to the $O(6)$ bosonic
sigma model in a suitable limit. In the $O(6)$ sigma model we have
a configuration with finite charge density whose free energy can
be computed as a solution of an integral equation \HasenfratzAB .
 The decoupling limit that gives   the $O(6)$ sigma model involves
 taking a  strong 't Hooft coupling and a
 small $J/(\log S)$ in such a way that
 quantum corrections remain finite.
Interestingly, the massive excitations of the $O(6)$ sigma model, which are in
the vector of SO(6),  can
be interpreted as insertions of the fundamental scalar  fields $\phi^I$.

This provides a way to compute higher loop corrections in the string
theory side without too much effort. These results could then  be
compared  to a suitable generalization of the BES equation \bes\ for
finite $J/(\log S)$, which will probably be sensitive to higher
order corrections of the phase of the fundamental magnon $S$ matrix.
 (We do not perform this computation here).

This paper is organized as follows. In section two we explain the $\log S$ scaling
of high spin operators using the symmetries of a conformal gauge  theory. We then
discuss the related issue of the double logarithmic infrared divergencies in the
Sudakov form factors. We also discuss how to simplify the strong coupling
computation of   $f(\lambda)$ in ${\cal N}=4$ super Yang Mills
at tree level and one loop by using the symmetries we mentioned above. (These computations  were
originally performed in \gkp\ and \FrolovQE ).

In section three we discuss high spin operators in ${\cal N}=4$ super Yang Mills and the
reduction to an O(6) sigma model.

Finally, in section four we present come conclusions.

\newsec{ High spin operators and  Sudakov factors in conformal gauge theories}

\subsec{High spin operators and the $\log S$ scaling }

In this section we would like to offer a geometric argument for the
logarithmic behavior of the anomalous dimensions of high spin
operators in gauge theories. Namely, we consider operators with very
high spin $S \to \infty$ keeping the twist finite. For simplicity,
consider first operators of the schematic form \eqn\quarb{ { \cal
O}_S = \bar q ( {\cal D}^{\leftrightarrow} )^S q } where $q$ is in
the fundamental representation. These operators have conformal
dimensions of the form \refs{\GrossCS,\KorchemskySI}
\eqn\confdim{
 \Delta - S = { f(\lambda) \over
2 }  \log S
}
 for large $S$. The factor of 1/2 is a convention.
 Here we are considering a theory with a
large number of colors and we are disregarding the mixing between
operators with different numbers of quarks. Alternatively, we could
be considering the weakly coupled theory in which this mixing is
suppressed in perturbation theory for the lowest twist operators.

We can also consider an operator of a similar form in a theory with
only adjoint fields, such as ${\cal N}=4$ super Yang Mills.
 In that case we consider a single trace operator of the schematic
form $Tr[ \phi^I ({\cal D}^{\leftrightarrow} )^S \phi^I ]$. In the planar limit the high
spin anomalous dimension  goes as $\Delta - S = f(\lambda) \log S
$, which is twice the value we had in \confdim\ because, at large
$N$ we can view an adjoint particle  a quark and an
antiquark, each of which gives rise to \confdim .

We now present an argument that explains the $\log S $ behavior in \confdim .
A previous argument can be found in \KorchemskySI .
In a conformal field theory, the anomalous dimension of an operator
is equal to the energy of the corresponding state of the field
theory on the cylinder $R \times S^3$. On the cylinder,
  a high spin
operator consists of two particles (or group of particles) that are
moving very rapidly along a great circle of $S^3$. These particles
are colored and the color field lines go between  the two particles.
See figure 1 .

\ifig\cylinder{Quark and antiquark moving very fast on opposite
sides of the cylinder. They become localized and can be replaced by light-like Wilson lines.
} {\epsfxsize1.5in\epsfbox{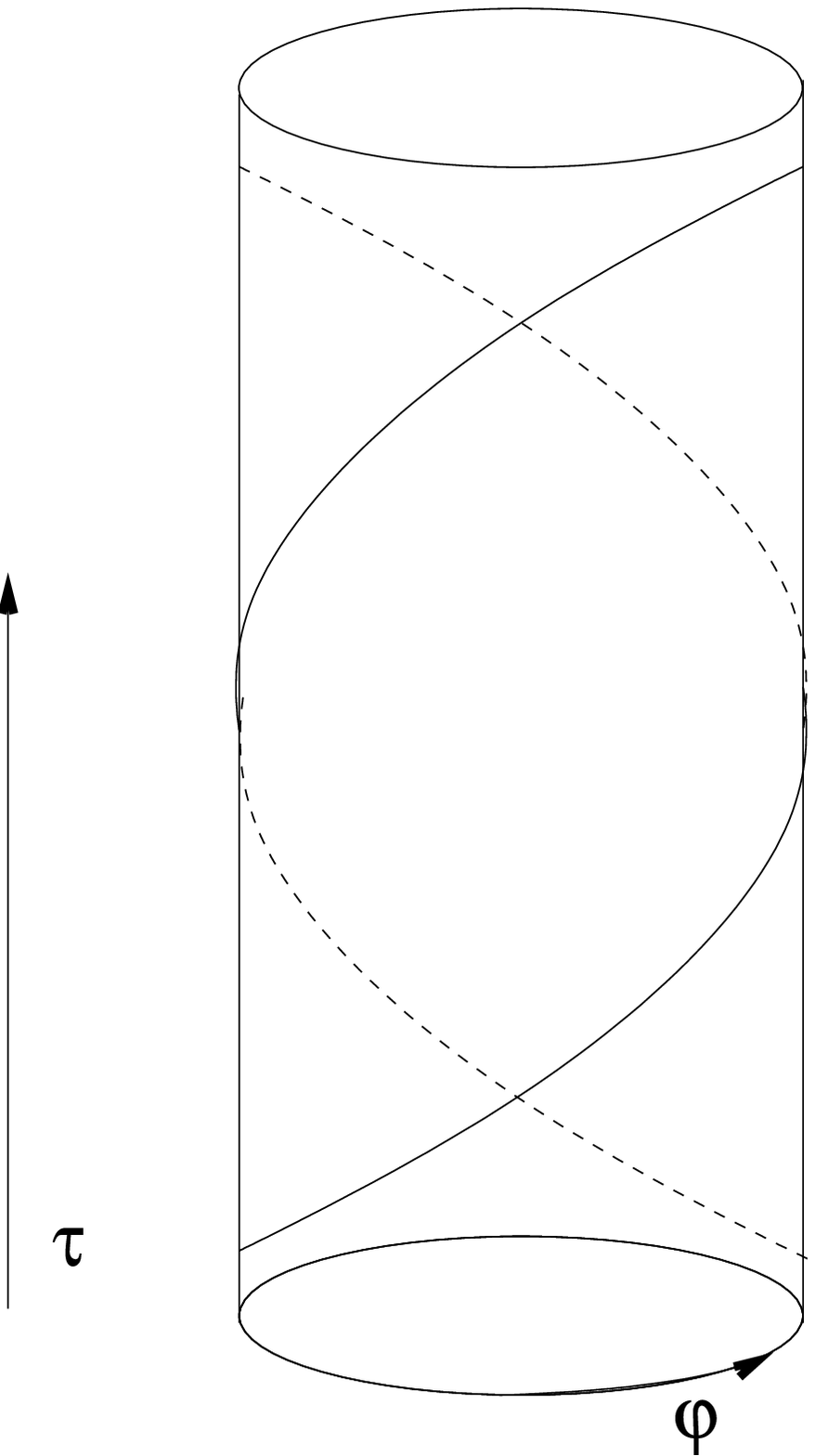}}

For simplicity, let us first consider the case of a quark and an
antiquark moving very fast along a great circle of the $S^3$, with
color gauge fields joining them. Parametrizing the cylinder as
\eqn\cylinder{ ds^2_{R\times S^3} = - d\tau^2 + \cos^2 \theta
d\varphi^2 + d\theta^2 + \sin^2 \theta d\psi^2 } we can then imagine
that the quark is close to the line $\varphi =\tau $ and the
antiquark at $\varphi = \tau + \pi$, both at $\theta=0$.

Notice that if we had a color neutral object that is moving fast on
the sphere, then its energy would go like $S$, namely $\Delta - S \sim
$finite, since we can get a particle which is moving fast along
the equator from a particle that is at rest by applying conformal
transformations. In other words from an operator ${\cal O}$, we can
consider its descendent $\partial^S {\cal O}$. In our case we have a
pair of particles and each particle carries color indices. In this
case we have a large contribution to $\Delta -S$ from  the
color electric field lines emanating from the particles. In order to
evaluate the effects of these fields, it is convenient to
 replace
 the quark and the anti-quark by a Wilson line. Thus, we first
  consider a Wilson line,
 which corresponds to $S = \infty$, and then we
 go back to  finite $S$.
We consider a lightlike Wilson line at $\theta=0$ and $\varphi =
\tau$ and an oppositely oriented line along  $\varphi = \tau + \pi$,
see figure 1 . This configuration is clearly invariant under $ \tau
\to \tau + c , ~\varphi \to \varphi + c$ where $c$ is a constant.
Less  obvious is the fact that these Wilson lines are  also invariant
under a second symmetry, which acts as a conformal transformation
which  is not an isometry of $R \times S^3$. As we will see, the
$\log S$ behavior is associated to this second symmetry. In order to
exhibit this symmetry
 more clearly  we can make a Weyl transformation of the $R \times S^3 $
 metric
\cylinder\ to $AdS_3 \times S^1 $ by writing \eqn\newparm{\eqalign{
ds^2_{R\times S^3}  = & \sin^2 \theta \left[  { - d\tau^2 + \cos^2
\theta d\varphi^2 + d\theta^2 \over   \sin^2 \theta} +  d\psi^2
\right] = \sin^2 \theta  ds^2_{AdS_3 \times S^1} \cr ds^2_{AdS_3} =
& - \cosh^2 \rho d\tau^2 + \sinh^2 \rho d\varphi^2 + d\rho^2
~,~~~~~~~~~ \sinh \rho = { 1 \over \tan \theta } }} A conformal
field theory should be Weyl invariant and thus we should get the
same result for the Wilson loop if we consider it on $AdS_3 \times
S_1$ \foot{ A CFT can have Weyl anomalies that are local and thus
should  not  affect the results for the non-local part of the Wilson
loop expectation value that we are considering.}.
 $AdS_3
\times S^1$  is the space where the field theory is defined and it
should not be confused with the $AdS_5$ space that will appear later
when we consider the gravity dual of the field theory. For the
moment we are making an argument purely in the context of the field
theory and is valid regardless of the value of the coupling or whether the
theory has a gravity dual or not.
 Thus we are considering the four dimensional field theory on
a four dimensional space which happens to be $AdS_3 \times S^1$. The
Wilson lines, which sit at $\theta=0$, are mapped to a pair of lines
along the boundary of $AdS_3$ at $\rho = \infty$.
 It is now convenient
to introduce new coordinates where the  $AdS_3$ metric takes the
form \eqn\newcoord{ ds^2_{AdS_3} = - du^2 + d\chi^2 - 2 \sinh 2
\sigma d u d\chi + d\sigma^2 } These coordinates arise by viewing
$AdS_3$ as the $SL(2,R)$ group manifold parametrized as \eqn\newme{
g = e^{ i u \sigma_2 } e^{ \sigma \sigma_3} e^{  \chi \sigma_1 } }
where $\sigma_i$ are the usual Pauli matrices. In contrast, to get
the metric in \newparm\ we set
  \eqn\globchoice{ g = e^{ i
\sigma_2 ( { \tau + \varphi \over 2} - { \pi \over 4}) }
 e^{ \rho \sigma_3} e^{ i \sigma_2
({ \tau - \varphi \over 2} + {\pi \over 4 } ) } }  The explicit relation between the two
coordinates is
\eqn\invrel{\eqalign{\sinh2\sigma=& - \sin(\tau-\varphi)\sinh 2\rho
\cr
e^{ 4 i u } = &  e^{2i(\tau+\varphi)}{\cos(\tau-\varphi)+i
\cosh 2\rho \sin(\tau-\varphi)\over \cos(\tau-\varphi)-i \cosh
2\rho \sin(\tau-\varphi)}
\cr
 \sinh 2\chi=&{\cos (\tau - \varphi) \sinh 2\rho \over
\sqrt{1+\sin^2(\tau-\varphi)\sinh^2 2\rho}}}}

One can see that in the new coordinates the Wilson loop is at $\chi
\to \pm \infty$ and at $\sigma =0$. Thus in the new coordinates
\newcoord, the two commuting
 non-compact symmetries of
the problem correspond to explicit isometries. Namely, to shifts in
$u$ and $\chi$.
  In particular, we have that the
Hamiltonian in the new coordinates corresponds to $\Delta - S = i
\partial_u$. Note that the $SL(2)_L\times SL(2)_R$ isometries of
$AdS_3$ act on $g$ as left and right matrix multiplication. The two
commuting isometries corresponding to $u$ and $\chi$ translations
are embedded in $SL(2)_L$ and $SL(2)_R$ respectively.

Since the Wilson loop is at the boundary, we end up with a
configuration where we have some color electric flux in the $u,\chi$
directions. It turns out that the flux is localized in the
direction $\sigma$ due to the warp factor in \newcoord. Thus the
flux leads to a constant energy density per unit $\chi$ and  the
energy is extensive in $\chi$.

Let us now explain in more detail why the energy is confined in the
direction $\sigma$. Note that for large $\sigma$ the $\sinh 2
\sigma$ term in \newcoord\ dominates. For very large and positive
$\sigma$ we have that $ds^2 \sim e^{2 \sigma} du d\chi$. Thus we can
view $u$ and $\chi$ as lightcone coordinates of a two dimensional
space with a large warp factor or gravitational potential. Thus the
flux is pushed towards smaller values of $\sigma$. For very
large and negative $\sigma$ we can make a similar argument. The
conclusion is that the flux is concentrated around $\sigma \sim
0$. Note that the direction $\psi$ in \newparm\ is
compact, so that the flux cannot dissipate in this direction either.
 In appendix A we consider explicitly the case of a $U(1)$ gauge
field and we show that indeed the flux is confined to the region
around $\sigma \sim 0$.
The computation in  appendix A also provides a
derivation for the one loop computation for the energy density and
$f(\lambda)$.

The conclusion of this discussion is that the expectation value of
the Wilson loop is divergent because of the infinite extent of the
$\chi $ direction, but it has a finite energy density per unit
distance in the $\chi$ direction, due to the fact that the flux does
not dissipate due to the gravitational potential in the $\sigma$
direction which leads to a confining potential.
 More precisely the energy, $\Delta - S$,
for the configuration is \eqn\energcon{ \Delta - S = { 1 \over 2}
f(\lambda) \Delta \chi } where the $1/2$ is simply a convention.

Now we would like to relate $\Delta \chi$ to $S$. First notice that the spin
is an isometry of $AdS_3$ in \newparm.
The spin generator written in terms the
coordinates in \newcoord\ has terms which go like $e^{ \pm 2 \chi
}$ (see the explicit expressions in  appendix A). In order to see
this,  note that $2 \chi$ translations are conjugate to the generator
${ \sigma_1 \over 2 }$ \newme , while the other $SL(2)_R$ generators
have charges plus or minus
 one under the action of this generator.
 This implies that we have exponentials of the form
 $e^{\pm 2 \chi}$. In the case that we have finite spin, we expect that the quark and
 the anti-quark are sitting around $\pm \chi_0$ respectively.
This would lead to a spin of the form
 $S \sim e^{2 |\chi_0|}$, or
 \eqn\chispin{
 \Delta \chi = 2\chi_0 =   \log S
 }
 Another way to say this is to note that the contribution to the spin from the flux alone
 goes
 as an integral over $\chi$ of a factor that grows like $e^{ 2 \chi}$. So, if the configuration
 has spin $S$ we need to cut off this integral around $e^{ 2 |\chi_0|} \sim S $.
Then for a quark-antiquark high spin operator we get
\eqn\qqbarexp{ \Delta - S = { 1 \over 2} f (\lambda) \log S }
while for a single trace operator made of adjoint fields we
get \eqn\adjloop{ \Delta - S = f(\lambda) \log S } Thus we see
that $f(\lambda)/2$ has the interpretation of the energy density
of the flux configuration along $u,\chi$ in the coordinates
\newcoord .

 We should also mention that one can consider an
operator containing $n$ fast moving partons. In the planar limit,
the flux  joins neighboring partons and we   have a
contribution going like $ \Delta -S = n { f \over 2 } \log S $ if
the spins of all partons are equal. This type of configurations
were studied in \refs{\manypartons,\kruczmany}.

\subsec{Finite $N$ and relation to two dimensional  QCD }

 Let us make now
some remarks about the finite $N$ case. If we have
dynamical quarks, or if we consider the adjoint case \adjloop\
then we see that we can nucleate colored particles that can
screen the flux. In that case the energy no longer scales like
$\log S$. Of course, this is  energetically convenient only once $
f(\lambda) \log S$ (or $f(\lambda) \Delta \chi $)
 becomes of order one. Thus, within the context
of perturbation theory, where $f(\lambda)$ is very small,  we can
ignore this issue and argue that we have the $\log S$ scaling also
for finite $N$ \foot{
 If
$N$ is finite the coefficient for the adjoint operator is not
equal to twice the coefficient for quark-anti-quark operator.}.
Note that the nucleation probability goes as $e^{ - { 2 \pi \over f(\lambda)}}$ and it is
very small as long as $\lambda $ is small.  
However, for strong enough coupling the leading twist operators
are not  single trace operators.   This is relevant for deep
inelastic scattering processes in strongly coupled field theories
\jpms . On the other hand,
 in ${\cal N}=4$ super Yang Mills, we can consider the lightlike Wilson loop operator
  for fundamental {\it external} quarks.
  In this case, the flux configuration is protected by a $Z_N$ symmetry, the center of
the gauge group. Thus in ${\cal N}=4$ super Yang mills we have a well defined problem
in terms of which we can define   $f(\lambda, N)/2$, for arbitrary values of the arguments.
  We can also
consider the strong coupling limit and relate this to a 't Hooft
loop. We can also consider the function $f$ for higher
representations. In the large $N$ limit the result is simply $n$
times the result for the fundamental, where $n$ is the number of
boxes and anti-boxes for the representation. On the other hand,
for finite $N$ the result  depends only on the $N$-ality (charge
under $Z_N$) of the representation. Such configurations were
considered at strong coupling in
  \ref\armoni{
  A.~Armoni,
  JHEP {\bf 0611}, 009 (2006)
  [arXiv:hep-th/0608026].
} using   D5 branes in $AdS_5 \times S^5$
which are very  similar to the ones appearing in the discussion of
1/2 BPS Wilson loops \ref\halfbps{
  N.~Drukker and B.~Fiol,
  JHEP {\bf 0502}, 010 (2005)
  [arXiv:hep-th/0501109].
  S.~Yamaguchi,
  JHEP {\bf 0605}, 037 (2006)
  [arXiv:hep-th/0603208].
  S.~A.~Hartnoll and S.~Prem Kumar,
  Phys.\ Rev.\  D {\bf 74}, 026001 (2006)
  [arXiv:hep-th/0603190].
   J.~Gomis and F.~Passerini,
  JHEP {\bf 0608}, 074 (2006)
  [arXiv:hep-th/0604007].
}.

In fact, it is interesting to consider the question of computing $f$ for different
representations at weak coupling. Once we view the problem in the coordinates we have
proposed\newparm , \newcoord , we see that we can do a Kaluza Klein reduction on to
the directions parametrized by $u$ and $\chi$. The $u$-energies
of various modes are given
by the values of $\Delta -S$. Thus we see that all modes in the field theory are
massive except for the gauge field along the $u,\chi$ direction. Thus we have
a reduction to a 2d QCD problem. This leads to an effective low energy action
\eqn\loweng{
S = - { 1 \over 4 g_2^2 } \int du d\chi Tr[F^2] + \cdots
}
where the dots indicate higher dimension operators we will describe below. The
 effective two dimensional coupling comes from integrating out all the Kaluza Klein
modes. We thus get
\eqn\effecu{
{1 \over g_2^2 } = { \pi^2 \over g_4^2} \left( 1 + (const) g_4^2 N + \cdots \right)
}
where $g_4$ is the four dimensional coupling. In
 principle this can include planar as well as non planar contributions\foot{
Note that if we start with a $U(N)$ theory we have to distinguish
between the $g_2$ for $SU(N)$ and the one for $U(1)$. }.

Notice that this effective field theory description is correct as long as there is
large separation between the Kaluza Klein scale which is of order 1 and the 2d QCD
scale which is of the order of $g_2^2 N $. This will be the case as long as we are
at weak coupling. If the 2d QCD lagrangian were the full description, then we would
conclude that for a general representation $R$ we have Casimir scaling for $f$, since
that is what we get in 2d QCD \ref\twodqcd{
A. Migdal, Zh. Eksp. Teor. Fiz. {\bf 69} , 810 (1975) (Sov. Phys. JETP. {\bf 42} 413); B.
Rusakov, Mod. Phys. Lett. {\bf A5}, 693 (1990);
 V. Kazakov, Zh. Eksp. Teor. Fiz. {\bf 85}, 1887 (1983) (Sov. Phys. JETP. 58 1096); I.
Kostov, Nucl. Phys. {\bf B265}, 223 (1986);  D.~J.~Gross and W.~I.~Taylor,
  Nucl.\ Phys.\  B {\bf 400}, 181 (1993)
  [arXiv:hep-th/9301068].
  }
\eqn\valef{
{f \over 2 } = { g_2^2 \over 2}  C_2(R)
}
However, this is not the full story. As we integrate out the massive fields, we can
get other operators beyond $F^2$. These operators were represented as dots in \loweng\ .
The first operators
 we can write down which are consistent with the symmetries of the problem are
\eqn\firstco{
 S = - { 1 \over 4 g_2^2 } \int du d\chi Tr[F^2]  + c N \int Tr[ F^4] + c' \int Tr[F^2]
 Tr[F^2]
 }
 where $c, ~c'$ are  numerical constants.
 Such   operators  lead to a violation of Casimir
 scaling at four loops. Thus we have the
 prediction that in any theory we would get Casimir scaling up to three loops, and then at
 four loops we will get a violation of Casimir scaling.
We also see that this effective field theory description
breaks down when the 2d QCD scale gets to
be of the order one, namely, when $g^2N \sim 1 $. In that case we should
consider the full theory.

\subsec{Conformal gauge theories in other dimensions}

The argument given here for the logarithmic scaling of
anomalous dimensions of high spin operators
 is completely geometrical and can also be
generalized to other dimensions were we can have conformal field
theories with a gauge symmetry. For a field theory in $D$ dimensions
we can do a Weyl transformation of the metric to write the metric as
$AdS_3 \times S^{D-3}$ and repeat the above arguments. In the case
of $D=3$ we get two copies of $AdS_3$ which are connected through
the boundary conditions for the fields. In the two dimensional case,
it might also be possible to find an argument (see \ref\BachasVJ{
  C.~Bachas, J.~de Boer, R.~Dijkgraaf and H.~Ooguri,
  JHEP {\bf 0206}, 027 (2002)
  [arXiv:hep-th/0111210].
}
 for a related problem)
 but we leave this for the future.

\subsec{High spin limit of double trace operators}

Notice that a crucial part of the argument leading to the logarithmic
scaling is the presence of a conserved flux. In cases where we do not have
a conserved flux we do not have a $\log S$ scaling.
As an example, we can consider
the behavior of double trace operators in a conformal theory, such as
${\cal N}=4$ super Yang Mills.
Here we consider operators of the schematic form
$ {\cal O}_d = {\cal O}_s ( {\cal D}^{\leftrightarrow})^S {\cal O}_s $, where
${\cal O}_s$ are gauge invariant single trace operators. To leading order in $N$
the dimension of these operators goes  like $\Delta - S = 2 \Delta_s$, where
$\Delta_s$ is the dimension of the single trace operator.
Here we consider the $1/N^2$ correction to this result.

The discussion we had above regarding the symmetries applies to this case too.
However, in this case we do not have a color flux along $\chi$. Instead we are
exchanging color neutral states between the states created by the
two operators ${\cal O}_s$.
Let us first study which states can be exchanged. Our first task is to
determine the energies of the possible states. The $u$-energies are simply given by
$\epsilon = \Delta-S$, where $\Delta$ is the energy of the state in the cylinder.
Thus we expect to find that we get a potential of the form
$V \sim e^{ - (\Delta_e - S_e) \Delta \chi }$, where $\Delta_e$, $S_e$ are the conformal
dimensions and spin of the exchanged particles\foot{In performing this argument we have
implicitly assumed that the operator that performs $\chi$ translations has eigenvalues
which are related to the ones of the operator performing $u$ translations. This
can be understood from analytic continuation in the metric \newcoord .}.
Using  \chispin\ we find that for large spin $S$ we get
\eqn\correct{
\Delta - S = 2 \Delta_s + { const} { 1 \over N^2 } { 1 \over S^{(\Delta_e - S_e ) }}
}
The leading power comes from the operator with the lowest value of $\Delta_e - S_e$. Notice
that $\Delta_e$ and $S_e$ are not large.
In general the possible exchanged states are subject to selection rules due to
the symmetries of the operator ${\cal O}_s$, thus we cannot simply take the operator with
lowest $\Delta_e -S_e$ in the spectrum of the theory.
However,  we can always exchange
the stress tensor operator, which has twist $\Delta_e - S_e =2$. Thus, the power of $S$ in
\correct\ is bounded by $1/S^2$. Notice that this result is valid whether or not the
gauge theory has an $AdS$ dual or not. Our argument is purely field theoretic and it is
valid for any conformal theory, including non-gauge theories.

The result \correct\ agrees with the more detailed analysis
performed in \refs{\costa,\costathree} which used the gravity description
 \foot{See section 4.6 in \costathree\ and take
$h  \sim S$, $\bar h \sim $ small,  $\Delta \to \Delta_e$, $j\to S_e$.}.
The more detailed analysis of \costathree\ makes it possible
 to consider more general operators, but the leading dependence
 on $S$ for large spin is fixed  by this argument\foot{Maybe is possible to make a
 general field theory argument for all the operators considered in \costathree . }.

As we will see below, these results also explain
  why do not get double log Sudakov divergences in theories without
conserved fluxes, such as the $\phi^3$ theory in six dimensions considered in
\sudakovreview .

In summary, the reason for the $\log S$ is simply that the
configuration develops  an additional symmetry in the infinite spin limit.
This symmetry becomes manifest when we do a Weyl rescaling of the
metric. We then see that the energy is extensive along the
coordinate conjugate to this symmetry generator. For finite spin, we
have only a finite range for this coordinate, a range proportional to $\log S$.

\subsec{Sudakov factors and the cusp anomalous dimension}

Another issue that can be understood by performing Weyl
transformations in the metric and coordinate redefinitions is the
behavior of soft divergences in scattering amplitudes.
It is well known that exclusive scattering amplitudes of colored (or charged)
have infrared divergencies due to the emission of low energy gluons (or photons).
These IR divergences disappear when we consider a physical observable (see \factorization\
for example), but are sometimes replaced by explicit dependence on detector resolutions or
parameters entering in the definition of jet observables. For this reason a great
deal of effort was devoted to understanding the structure of these divergences. These
divergences can be resumed into  an expression of the form
\refs{\sudakovreferences,\Korchsudakov,\sudakovreview,\msterman,\catani,\bds}
\eqn\sudfa{
{\cal A} \sim e^{ - h(\lambda) (\log \mu_{IR})^2 - h'(\lambda) \log \mu_{IR} }
}
where $h$ is some function of the coupling.
In a planar gauge theory the color of each gluon is correlated with the anticolor of
the next and so on. For each consecutive pair we get a factor of the form \sudfa , with
a function $h$ given by $f/4$.
These double logarithmic divergences
can be
computed by replacing the hard gluons by Wilson loops \Korchsudakov\ (see also
\sceft\ for a more systematic discussion). In
particular, we have light-like Wilson loops with a cusp. We would
like to see that this cuspy Wilson line in the fundamental representation
has a behavior of the form
\eqn\cuspwline{ \langle W \rangle \sim e^{ -   { f(\lambda) \over 4
}  ( \log \mu_{IR}/\mu_{UV} )^2} } where $\mu_{IR}$ and $\mu_{UV}$
are the  UV and IR cutoffs. This  behavior depends on how we introduce
the UV cutoff. Here we are introducing it in the way that arises
when we consider scattering amplitudes, where the UV cutoff appears
when we can no longer replace the hard particles by Wilson lines\foot{If we were to choose
a boost invariant UV and IR regulator we would get a divergence due to the boost invariance.
The UV regulator we choose here is not boost invariant. }. In other words, imagine we
have two gluons coming out of an interaction region around the origin
with momenta $(k_-,0)$ and $(0,k_+)$. We can then replace the gluon
with momentum $k_-$ with a Wilson line along $x^+= t+x>0$ and a fuzziness in the direction
$x^-$ given by $\Delta x^- \sim 1/|k_-|$. the other gluon gives rise to a Wilson line
along $x^-=t-x>0$.

In this case we can start with the coordinates
 \eqn\coordsta{ ds^2 =ds^2_{AdS_3 \times S^1} =
{ - dt^2 + dx^2 + dr^2 \over r^2 } + d\psi^2 = { ds^2_{R^{1,3} }
\over r^2 } }
We can now choose the coordinates
\eqn\newcoordt{\eqalign{ { t \pm x \over r } =& \sin \alpha   e^{\pm
\gamma } ~,~~~~~~{ 1 \over r} = \cos \alpha e^{-\tau} \cr
ds^2_{AdS_3} = & - d \alpha^2 + \sin^2 \alpha d\gamma^2 + \cos^2
\alpha d\tau^2 }} In this case the Wilson loop sits at $\tau
=-\infty$ and $\chi \rightarrow \pm \infty$.
These coordinates cover only a portion of $AdS_3$, namely the region
$ 0 < t^2 -x^2 < r^2$. We can cover other regions by going through
$\alpha \sim 0$ and $\alpha =\pi/2$ which are only coordinate singularities, but
are otherwise smooth surfaces. For example, if we set $\alpha \to i \alpha'$, we go
to the region $t^2 -x^2 <0$ and with $\alpha \to i \alpha'' + \pi/2$ we go to the region
$ t^2-x^2-r^2 >0$ which is the  forward light cone of the point $t=x=r=0$ in $AdS_3$.

The region parametrized by \newcoordt\ intersects the boundary
along $t=\pm x >0$, which coincides
with the cuspy Wilson loop we want to consider.
 This is a time dependent background and we  consider first
 the ordinary $AdS_3$  vacuum
which   leads to a particular state at both the future and past
horizons in
\newcoordt . Then, starting and
ending with such a vacuum, we insert a Wilson line  operator at $\alpha =
\pi/4$, $\tau =-\infty$ and $\chi \rightarrow \pm \infty$. We
then have flux lines joining these two asymptotic  regions.
We now note that we can get the metric in \newcoordt\ by performing an
analytic continuation of the metric in \newcoord\
 \eqn\analycont{
 \alpha = i \sigma+\pi/4, ~~~~~~\tau=i u-\chi,~~~~~
 \gamma=i u+\chi
}
 Notice that the particular
point $\sigma=0$ corresponds to $\alpha=\pi/4$.
This implies that the flux configuration that appears in the Sudakov computation
leads to a factor
 \eqn\exponsup{ {\cal
A} = e^{ - { f(\lambda) \over 4} \int d\tau d \gamma   } } where $f$
is the same as the function that appeared  above \energcon .
 The factor of four  in
\exponsup\ is due to the change in measure
$\int du d\chi \to - i \int d\tau d \gamma $.
  This analytic
continuation is also responsible for the fact that we get a real
exponent in \exponsup . Thus, this analytic continuation explains
the connection between the $f$ that appears in the Sudakov factor
and the function $f$ that appears for high spin operators \Korchsudakov .
A similar analytic continuation was also used in \ultimokr\ to argue, from the
AdS side, that
the same function appears on both calculations.

Now we
need to relate the range of $\tau$ and $\gamma$ in \exponsup\ to the
IR and UV cutoffs. We expect the IR cutoff to be boost invariant so
that we get $ \tau < - \log \mu_{IR} $. If we want to relate this
computation to the computation of gluon scattering amplitudes, then
the UV cutoff is not boost invariant since the approximation of
replacing a hard gluon by a Wilson line   fails if we do a very
high boost since the gluon   ceases to be hard.
 Thus, we
get a constraint of the form \eqn\uvcutof{
 \tau_{UV} \equiv -\log \mu_{UV} < \tau \pm \gamma
 }
 In the application to gluon scattering the UV cutoff is
 related to the momentum transfer
between the two  gluons generating the flux,  $\mu^2_{UV} \sim -s $ \foot{
We have the cutoffs $\Delta x^\pm \sim 1/|k_\pm|$. Since the final answer
will be boost invariant we can take
 $\Delta x^\pm \sim 1/\mu_{UV}$, with $\mu_{UV}^2 = - s$.}. Thus we are
integrating over a triangular shaped region. The regions near each
boundary in \uvcutof\ correspond to the collinear regions. Near these
boundaries we can replace only one of the gluons by a Wilson line
but not both of them. Thus, we are integrating over a domain of the
form \eqn\domainint{ \eqalign{
& \int d\tau d\gamma = \int_{ \tau_{UV}}^{\tau_{IR}}  d\tau \int_{ -
(\tau -\tau_{UV} )}^{\tau - \tau_{UV} } d\gamma  = \left[ \log { \mu_{IR}\over \mu_{UV}} \right]^2
\cr
{\cal A} &\sim \exp\left\{  - {f \over 4 } \left[ \log  { \mu_{IR}\over \mu_{UV}}
\right]^2
\right\}~,~~~~
 ~~~~~
\tau_{UV} = - \log \mu_{UV} ~,~~~~~~~~~\tau_{IR} = - \log \mu_{IR}
}}
where we have neglected single log terms that can arise from the boundaries in \uvcutof .
By scale invariance, such region are expected to contribute with a term linear in $\tau$.
Thus we   get an additional term of the form \msterman\
\eqn\integr{ \exp \left\{ { 1 \over 2}  g( \lambda ) \int_{\tau_{UV}}^{\tau_{IR}
}  d\tau  \right\}
}

We should also note that using similar reasoning, but now considering the matching between
different regions of $AdS_3$ in \coordsta\ that we discussed above, we could consider a
non-lightlike cusp Wilson line with a boost angle $\gamma$. For a large angle $\gamma$,
then we expect an expression of the form
\eqn\anglela{
\langle W \rangle \sim e^{ - { f \over 4 } \Delta \chi \int d\tau } \sim
 e^{ - { f \over 4 } \gamma  \log { \mu_{UV} \over \mu_{IR} } } ~,~~~~\Delta \chi = \gamma
}
Which is the well known result \Korchsudakov , that the cusp anomalous dimension of a
non-lightlike Wilson loop \cusppolyakov , is linear in $\gamma$ for large $\gamma$.
The reader should not be confused by the fact that one calls
  $f$ a ``cusp anomalous dimension'', though
the scaling with the cutoff is different for a light-like cusp \cuspwline\ than for the
original, non-lightlike, cusp discussed in \cusppolyakov , which has a single log.

\subsec{Non-conformal theories}

It is also possible to understand the above behavior in
non-conformal theories.
 Here the new element is that the gauge theory has a scale, which
we   parametrize with $\Lambda$. We can think of this scale as
the UV cutoff, where we set the coupling constant. The coupling can
then be evolved to a lower scale via the RG equation. Then, as we do
a Weyl transformations of the metric by some function $\Omega(x)$,
$d{s}^2 = \Omega^2(x) d{s'}^2 $ we  find that the new theory in the
new metric, $ds'^2$,
 has a scale $\Lambda' = \Lambda \Omega(x) $. This means
that in the theory on the new space $d{s'}^2$ we are setting the
coupling to a constant at the scale $\Lambda'$ which depends on the
position. If we now were to choose a constant cutoff $\Lambda''$ in
the new theory we would find that the coupling on the new cutoff
scale $\Lambda''$ is not constant but $x$ dependent. This dependence
can be obtained by solving the RG equation to relate the constant
coupling at scale $\Lambda'(x)$ to the scale $\Lambda''$. As long as
the starting $\Lambda$ is sufficiently large, the value of
$\Lambda'$ is  large enough so that the dependence on $x$ is be
very slow and we can use the ordinary RG equation for a constant
coupling. In appendix B we discuss more explicitly the simpler case
where we map a non-conformal theory between the plane, $R^4$, and
the cylinder $R \times S^3$.

 In our
 case, in \coordsta\ we find that $\Omega \sim r$ and thus
the new scale is \eqn\newsca{ \Lambda' = \Lambda r=
\Lambda { e^\tau \over  \cos \alpha }
 }
 The $\alpha$ dependence is not too important for us. The
$\tau$ dependence implies
 that the Hamiltonian generating $\tau$ translations is $\tau$
  dependent and the value
 of the Wilson loop involves an integral over $\tau$. When we do that integral, it is
 important to remember that the range of $\gamma$ also depends on $\tau$ and the UV and IR
 cutoffs. Thus, when we repeat the above computations we   find that $g$ and $f$ are
 the $\tau$ dependent eigenvalues of the $\tau$-Hamiltonian. They depend on the coupling,
 which itself depends on $\tau$ through the $\beta$ function equation. If we use dimensional
 regularization in order to cutoff the IR divergencies, then we should use the $\beta$ function
 in $4 + \epsilon $ dimensions to determine the $\tau$ dependence of the coupling constant.
The bottom line is that we obtain an expression similar to the above
one but the function $f(\lambda(\tau)   )$ depend explicitly on the
``time" $\tau$ and should placed inside the integrals.  These
expressions agree with the expressions in \refs{\msterman,\bds}.

If we concentrate on the leading dependence on the IR cutoff, then
we are integrating in $\tau$ up to $\tau = - \log \mu_{IR}$ and for
each value of $\tau$ the range of $\gamma$ is $\Delta \gamma \sim 2
(  \log \mu_{UV} -\tau ) $.
  Thus we have
\eqn\newf{ \langle W \rangle \sim e^{ - \int_{-\log \mu_{UV}}^{ -\log \mu_{IR} }
 d\tau { f( \lambda(\tau) )  \over 4 } \Delta \chi(\tau) }
 \sim e^{ - \int_{-\log \mu_{UV}}^{ - \log \mu_{IR} }  d\tau
 { f( \lambda(\tau) )  \over 4 } 2 ( \log \mu_{UV} -\tau) }
} This agrees with the general expressions derived in \refs{\msterman,\bds}, for the
leading IR divergence of gluon scattering amplitudes.

Using the $AdS/CFT$ one can consider the computation of Wilson loops
with cusps in non-conformal theories. This was done
 for the Klebanov-Strassler cascading  theory in \cuspcascading and in
  $ 4 + \epsilon$ dimensions in
\aldaymaldacena .

\subsec{ High spin operators at strong coupling }

We now consider ${\cal N}=4$ super Yang Mills at strong coupling and analyze it using
the gravitational dual.
  From our
general discussion we concluded that $f(\lambda) $
 can be computed in terms of a light-like Wilson loop.
It is convenient to slice $AdS_5$ in coordinates where the boundary
is manifestly $AdS_3 \times S^1$
\eqn\writad{\eqalign{
ds^2 &= \cosh^2 \zeta ds^2_{AdS_3} + \sinh^2 \zeta d\psi^2 + d\zeta^2=
\cr
& =  \cosh^2 \zeta \left[  - du^2 + d\chi^2 - 2 \sinh 2
\sigma d u d\chi + d\sigma^2 \right] + \sinh^2 \zeta d\psi^2 + d\zeta^2
}}
These coordinates cover all of
 $AdS_5$ and  the boundary sits  at $\zeta \to \infty $.

In the gauge theory we considered a configuration with flux in the $u\chi$ direction
\newcoord . This
 gives rise to a string extended along the $u\chi$ directions of $AdS_3$. The warp
factor in the $\zeta$ direction pushes the string to $\zeta =0$, which is a $U(1)_\psi$ symmetric
point. In addition the warp factor in the $\sigma$ direction pushes the string to $\sigma =0$.

There are some interesting features of this string. First,
its tension gives us the energy
density. Thus, simply the tension of the string gives us the strong coupling behavior for
the cusp \refs{\gkp,\kruczwil}
\eqn\tenstr{
{ f \over 2 } = { T \over 2 \pi \alpha'} = { R^2 \over 2 \pi \alpha'} = { \sqrt{ \lambda }
\over 2 \pi }
}
We see that we get the result in a direct way without solving any equations.

Second, we can easily consider small fluctuations around this string configuration \FrolovQE . We
can see that for quadratic fluctuations we have a boost symmetry in the $u,\chi$ directions.
This is not a symmetry of the full problem, but it is a symmetry of the theory at the
quadratic level.  We can
easily find the bosonic excitations and we can compute their masses.  We find that there are
five massless goldstone
bosons associated to the motion on $S^5$. The oscillations in the $\sigma$
direction are described by a   massive goldstone field
with mass $m^2 =4$ that comes from the $SL(2)_L$ symmetries that the string breaks\foot{
Goldstone bosons can be massive when the broken symmetry does not commute with the Hamiltonian.}.
In other words, the creation and annihilation operators for the modes of
the $\sigma$ field on the worldsheet
come from $J_L^\pm$ in $SL(2)_L$,
recall that $2 J^3_L = i\partial_u$ is the energy.
This corresponds to   oscillations  in the $\sigma$ direction inside $AdS_3$.
Then there are two bosons of  $m^2 =2$
associated to motion in $\zeta e^{i \psi}$, these are not obviously Goldstone bosons.
Nevertheless one can view them as Goldstone bosons according the following heuristic
argument.   The full theory
has conformal symmetries which are not isometries of $AdS_3 \times S^1$. In particular
we have conformal generators in the spin $({ 1 \over 2}, { 1 \over 2},\pm 1)$ under
$SL(2)_L \times SL(2)_R \times U(1)_{S^1}$. These symmetries are broken by the string.
They create modes with wavefunctions of the form $e^{ - i u \pm \chi }$. We see that these
are exponentially growing in the $\chi$ direction and thus would carry momentum $p= \pm i$.
Thus, the dispersion relation should be such that $\epsilon (p=\pm i )=1$. If the dispersion
relation is relativistic, then we get $m^2=2$. This argument is heuristic because we are
talking about non-normalizable modes. Now let us turn to the fermions.
  All the fermions have the same mass since
 they have to transform under the spinor representation of $SO(6)$ and the lowest dimensional
 representations have
  four complex dimensions, corresponding to eight real fermions.
 They all have $m=1$ which can be viewed again as goldstone fermions. Their mass is fixed
 by the transformation properties
 of the supercharges under $SL(2)_L$, where $SL(2)_L$ are the left isometries
 of $AdS_3$.
The advantage of viewing them as Goldstone bosons or fermions is that their energies   stay
with these values (at zero momentum) as long as the symmetry is not restored\foot{
We will  see that the 5 massless modes will actually get a mass non-perturbatively in
the $\alpha'$ expansion and the SO(6) symmetry will be restored.}.
>From this spectrum of masses it is straightforward to compute the vacuum energy and we obtain
the one loop contribution \FrolovQE
\eqn\vacuumen{
{ f_1 \over 2} = \int_{-\infty}^\infty   { dp \over 2 \pi } { 1 \over 2}
 \left[ 5 |p| + \sqrt{ p^2 + 4 } + 2 \sqrt{ p^2 + 2} - 8 \sqrt{p^2 +1 } \right] = - { 3 \log 2
 \over 2 \pi }
 }

 One can also check explicitly that this string configuration breaks all the supersymmetries.
 Thus it is different from the string configuration corresponding to the circular Wilson loop
 which is BPS.

 \newsec{The O(6) sigma model from string theory}

In this section we consider further the worldsheet theory describing the string
associated with highly spinning operators or lightlike Wilson loops.
We consider a string stretched along the $u,\chi$ coordinates in  \writad , and
we work in static gauge.
Notice that the theory is not invariant under boosts in the $\chi$
and $u$ directions, but it does become boost invariant at low
energies. In fact, the spectrum we discussed above is precisely
boost invariant. Let us now imagine taking a low energy limit where
we look at the system at distances (in $\chi$ ) much larger than
one, which is the mass of the fermions and the order of magnitude of the mass
of the massive bosons.
 In this case
only the massless excitations survive. In two dimensions, massless
fields have large fluctuations which lead to interesting dynamics in
the IR. In our case, the massless fields describe an $S^5$. In other
words, they describe the $O(6)$ sigma model. This is a model where
the coupling becomes strong in the IR and the theory develops a mass
gap. Moreover, this is an exactly solvable theory \zamolzamol . The
scale set by the mass of the fermions (and massive bosons) acts as a
UV cutoff for the O(6) sigma model, where the O(6) theory merges into
the full $AdS_5 \times S^5 $ sigma model.

When we compute  the cusp anomalous dimension at strong coupling we
are computing the vacuum energy of this theory. Of course, the
vacuum energy in the $O(6)$ sigma model is UV divergent. In our
case, this UV divergence is cut off at the scale where the
fermions start contributing. We can see this explicitly in the
one loop result \vacuumen\ . Thus the vacuum energy does not seem to
have a clear contribution that comes purely from the $O(6)$
sigma model. A two loop computation was attempted in \ref\twoloopstr{
 R.~Roiban, A.~Tirziu and A.~A.~Tseytlin,
  arXiv:0704.3638 [hep-th].
}.

There are some interesting
features of this relation to the $O(6)$ model. First, note that the
massive excitations of the theory transform in the vector
representation of $O(6)$, thus it is natural to identify them with
the fundamental scalars $\phi^I$ of the gauge theory. It looks like
such excitations should appear naturally as we compute the spectrum
of charged operators around the lowest twist high spin  operator.

This relation to the O(6) sigma model   allows us to perform all
orders computations in $AdS$ by focusing on the right observable. We
choose an observable which receives most of its contribution from
the low energy region described by the $O(6)$ sigma model. There is
an interesting concrete set of operators that has been studied
recently \refs{\FrolovQE,\BelitskyEN}
 which has a limit that can be explored in terms
of the $O(6)$ theory. These are operators which carry large spin $S$
and also large charge $J$, where $J$ scales like $\log S $. In other
words, consider single trace operators in planar ${\cal N}=4$ super
Yang Mills
 with
\eqn\newj{ S, ~J \to \infty ~,~~~~~~~~~ j \equiv { J \over 2 \log S
} = {\rm fixed } }
 (the factor of 2 is for convenience). Using the arguments in the previous section, which
 connect the spin $S$ to the extension of the string (or corresponding field theory
 configuration)  in the $\chi$ direction we see that in
 this limit we have a configuration with finite current density along the $\chi$
 direction on the worldsheet.
 We then conclude that the anomalous dimensions scale as $\log S$ and that
 \eqn\anomgo{
\lim_{S \to \infty}  {  \Delta - S \over \log S}  =  f(\lambda ) +
2 \epsilon(\lambda, j)
 }
where $f(\lambda)$ is the cusp anomalous dimension and $2
\epsilon(\lambda, j)$ is the additional energy due to the SO(6) charge
density $j$. Note that $\epsilon(j=0)=0$. The argument of the previous section
implies that this is
the right scaling for all values of $\lambda$ and $j$.

Note that at strong coupling we can consider the same string we
discussed above, which is stretched along the $u$ and $\chi$
directions. The string carries a current density proportional to $j$
since $2 \log S = \Delta \chi$ is the length of the folded string
corresponding to a single trace operator of spin $S$.  Similarly the factor of 2 in
\anomgo\ is chosen so that $\epsilon(j)$ is the energy density along a single string
stretched along the $\chi$ direction.

We can compute $\epsilon(j)$ using the
 $O(6)$ sigma model in the regime
where the characteristic time variations of the
angular coordinates are much smaller than the mass of the
fermions. We have \eqn\curreden{ j = { \sqrt{ \lambda } \over 2
\pi } \dot \varphi }  We require $\dot \varphi
\ll 1 $, thus we want $j \ll \sqrt{\lambda }$ for starting to
trust the $O(6)$ results. The classical sigma model result is
\eqn\classans{ \epsilon(j) = { \sqrt{\lambda} \over 4 \pi
} \dot \varphi^2 = { \pi \over \sqrt{\lambda} } j^2 }
The strong infrared dynamics of the O(6) sigma model   generates a
mass gap \eqn\massexc{ m = k \lambda^{1/8} e^{ -{1 \over 4}
\sqrt{\lambda} } [ 1 + o( { 1 \over \sqrt{\lambda} } ) ] } where
$k$ is a constant that depends on the details of how the $O(6)$
model is embedded in the full $AdS_5 \times S^5$ string sigma
model.  This formula is valid at large
$\sqrt{\lambda}\gg 1$ so that the $O(6)$ sigma model can be
suitably decoupled from the rest.

Let us specify more precisely the decoupling limit that gives the O(6) model.
 We   take the limit $S \to
\infty$ with $j$ fixed \anomgo . We then take the limit
\eqn\decoupl{
\lambda \to
\infty ~, ~~~j\to 0 ~,~~~~~~{\rm with}  ~~~~~~~
 {j \over m}  = j k^{-1} \lambda^{- 1/8} e^{ {1 \over 4}
\sqrt{\lambda} } = {\rm fixed}
}
In this limit we find that \eqn\findosix{ \epsilon(j) = j^2 {\cal
E}(j/m) } where we used dimensional analysis. The function ${\cal
E}$   can be determined purely in terms of the $O(6)$ sigma model
\HasenfratzAB . For large $j/m$ this function can also be expanded
using $O(6)$ perturbation theory. Thus, we can use the $O(6)$
results to compute the $\alpha'$ expansion of this observable.

The problem of computing the energy of a configuration with constant
current density was considered in \HasenfratzAB . These authors
derived an integral equation determining the energy as a function of
the chemical potential $h$ for the charge $j$. They found that
\eqn\freeen{ F(h) = h^2 {\cal F} (h/m)=-{h^2 \over 2}\left(-\beta_1
\log(h/m)+{\beta_2 \over \beta_1} \log\log(h/m)+c+\tilde{c}
{\log\log(h/m) \over \log(h/m)}+... \right) } where the coefficients
$\beta_1$ and $\beta_2$ can be related to the one and two loop beta
function coefficients in the O(6) theory.  Of course the first two
terms can also be easily computed using perturbation theory in the
sigma model. But the computation of $c$ amounts to a computation of
the mass gap which is more involved \HasenfratzAB . In  appendix C
we review \HasenfratzAB\ and give  the values of these coefficients
for an O(N) model. In  appendix C we also show that the structure of
the perturbative series can be used to fix the coefficients of the
logarithmic terms
 (for example $\tilde c$).  The energy density
discussed above can be computed by first computing $j$ from
\freeen\ and then performing the Legendre transform to get
$\epsilon(j) = F(h) + j h $. In this fashion we can obtain the
function ${\cal E}$ in \findosix .

This gives a precise prediction that could be used to test the
BES/BHL \refs{\bhl,\bes} guess for the S-matrix to all orders in the $\alpha'$
expansion, and probably in an exact way. Hopefully, all that one
needs to perform this comparison is to have a closer look at the
Bethe equations in the SL(2) subsector and repeat the steps in
\refs{\es,\bes,\kris} keeping $J/\log S$ constant.

 \subsec{$O(6)$ free
energy and comparison to string theory}

One loop string theory computations on a related regime have been
considered by  Frolov, Tirziu and Tseytlin in  \FrolovQE . They
considered closed string
configurations with
\eqn\defx{
 y \equiv {2 \pi \over \sqrt{\lambda} } j
 }
fixed\foot{ They wrote their results in terms of $x$, with  $x=1/y$.}.
As a check of what we have been saying we  show that
the limit of small $j$, $y$   of the formulas in
\FrolovQE , match precisely the expectation from the O(6) side.
In addition, using their computation we can fix the coefficient $k$ in \massexc , which
is sensitive to threshold corrections.

In the small $j$ limit their tree level and one loop results read
\eqn\xexpr{\eqalign{  \epsilon(j)  = { \Delta - S - f(\lambda) \log S \over 2 \log S } =  {y^2} \left(
{ \sqrt{ \lambda } \over 4 \pi }   - { 1 \over \pi} \log y +{ 3 \over 4 \pi }   \right)  }}
The tree level result is simply \classans .

For large values of $j$, $j/m \gg 1$ (but still with $j\ll
\sqrt{\lambda}$, so that the $O(6)$ description is valid), and using
the relations \massexc , \freeen\ and  \defx , we can
expand the energy density in powers of ${ 1 \over \sqrt{\lambda}} $.
We find
\eqn\lambdaexp{\eqalign{\epsilon(\lambda,y)_{0+1}=& {y^2 } \left[ {\sqrt{\lambda} \over
4 \pi }+ { \beta_1 \over 2 }  \log(y) + k_1
  \right] ~,~~~~~~~~ 2 k_1 = - c - \beta_1 \log k + { \beta_2 \over \beta_1} \log(-2 \pi \beta_1)
}}
Let us focus first on the
 $\log y $ term whose coefficient
 depends on $\beta_1$. We see that for the correct O(6) value, $\beta_1 = -2/\pi $,
 we obtain agreement with
 \xexpr\  \FrolovQE . The constant piece, $k_1$, depends on
 $k$ in \massexc . This is a quantity that we cannot determine by purely O(6) sigma model
 computations, since it depends on threshold corrections that involve the other massive
 modes. By matching the constant piece in the one loop answer in the full theory,
 computed in \FrolovQE\  (and reproduced in \xexpr),
  we can actually
 determine
\eqn\kvalue{k={2^{1/4} \over \Gamma(5/4)}}

Note that the computation in \FrolovQE\ was done keeping $ y$ fixed and arbitrary, while here
we are just focusing on the region $y\ll 1$ where the results can be computed using the $O(6)$ theory.
Indeed, the coefficient of the $\log y $ was
determined by $O(6)$ but not the constant part.

\subsec{Higher loop predictions}

Here we consider higher loop predictions for the energy density.
At higher loops and very small $y$,  the most important terms in the
expansion are the logarithmic terms. Such terms are determined by the
renormalization group
 equations in terms of lower order terms. The expression \freeen\ is the expansion of the free
 energy for the first two orders in perturbation theory. This expression also determines the
 $\beta$ function at one and two loops. We thus expect that using the renormalization group
 we could determine the two leading logs at each order in perturbation theory.
Of course, we could just blindly expand the expression \freeen\ and directly see that the
  two leading  logs are determined.
  Of course, these logs are renormalizing the coupling from the
UV scale where it is given in terms of $\sqrt{\lambda}$ down to the
scale set by $y$. However, since the coupling constant expansion in
the full sigma model is $1/\sqrt{\lambda}$ we might still wish to do
an expansion in powers of $1/\sqrt{\lambda}$. In that case we can
give the expression for the first two leading log corrections at
each order (after we used \kvalue ), ($\epsilon = \sum_{k=0}^\infty
\epsilon_k $)
%
%
%
\eqn\twoleading{\eqalign{
\epsilon(y)_{2}=&  {y^2 }
{1 \over \sqrt{\lambda} \pi
 }{\left(4  \log^2 y - 3 \log y  + \cdots \right)}
 \cr
 \epsilon(y)_{3}=& { y^2 } {1 \over \pi
\lambda}\left(-16 \log^3 y+ 6  \log^2 y  + \cdots \right)
\cr
\epsilon(y)_{l+1} =
 & y^2 { 1\over \pi \lambda^{l/2} } \left( (-1)^{l+1} 4^l \log^{l+1} y + (-1)^l
 4^{l-1}3
[ (l+1)(1 -h(l)) +  l ]\log^l y  + \cdots \right)
  }}
where $h(l) = \sum_{n=1}^l 1/n$ is a harmonic sum.

It would be nice to see whether these terms, which are easily computed, contain any information
about the higher order corrections for the dressing phase  \refs{\afs,\bhl,\bes}
when one computes this energy
using a suitable generalization of the techniques in \refs{\es,\bes,\kris}.
 If that is the case, then one
could test the higher order terms in the dressing phase. If the leading logs do not
depend on the higher order corrections to the phase, then one would be forced to consider
higher order corrections in the O(6) sigma model, which are still much easier to compute
than higher order corrections in the full $AdS_5 \times S^5$ string sigma model.

We should note that at each order in the $1/\sqrt{\lambda}$ expansion there are also terms
  which arise from higher order threshold corrections in \massexc . Such
terms are not calculable in the purely $O(6)$ theory. These higher order corrections disappear
if we take the decoupling limit \decoupl . Only the constant $k$ in \decoupl\ appears, and we
have already fixed it in \kvalue\  using the results in \FrolovQE .

\subsec{Very small $j$ limit }

Note that our discussion makes sense even non-perturbatively in the $O(6)$ coupling.
Thus, we can consider extremely small values of $j$.
In this regime we have a very low charge density, so we have well separated massive
particles in the $O(6)$ theory thus make the simple prediction
that \eqn\massen{ \epsilon(j) \sim  m j ~,~~~~~~~~~~j \ll m }
where $m$ is given in \massexc .
This
is valid for $\lambda \gg 1$.
Thus we have particles that transform in the vector representation of $O(6)$.
This is reminiscent of what happens at weak coupling, $\lambda \ll 1$, where for
low $j$ we also have an answer linear in $j$,
  $\epsilon(j ) \sim j$. This is simply the statement that the scalar fields $\phi^I$, which
  carry the $SO(6)$ charge contribute one unit to the twist ($\Delta-S$).
  We see that, for very small $j$, the
functional dependence on $j$ is the same at weak and strong coupling.
However, the coefficient is very
different. The small value of the coefficient at strong coupling signals the existence
of the region where
the physics is described by the O(6) sigma model, since there is large difference between the
mass gap of the $O(6)$ particles an the mass of the fermions, which sets the scale where the
$O(6)$ theory breaks down. At weak coupling, $\lambda \ll 1$,
 there is no such large separation and it would
be wrong to use $O(6)$ formulas to compute the energy.

It would be interesting to see if one can get similar reductions to an O(4) or O(3) sigma
models by considering strings in $AdS_3\times S^3$ or $AdS_3 \times S^2$.

\newsec{Conclusions}

In this article we have presented a simple picture for the cusp
anomalous dimension. This is a quantity that appears in various
computations in gauge theories. We found it convenient
to perform a Weyl transformation of the metric from $R^{1,3}$ to $AdS_3 \times S^1$,
which simplifies the action of the symmetries that determine the form of the results.
The cusp anomalous dimension   becomes the
energy density of a certain flux configuration of the gauge theory
on $AdS_3 \times S^1$. It is a flux configuration that is invariant under
two non-compact translation symmetries.   These
symmetries explain the logarithmic behavior of certain quantities.
For example, the logarithmic behavior of the dimension of high
spin operators, $ \Delta - S = f(\lambda) \log S$,
 arises when the configuration has a finite extent along
 the coordinate conjugate to one of the translation symmetries.
  Similarly, the flux configuration associated to the
Sudakov factor arises after performing an analytic continuation
and the double logarithmic behavior arises from imposing a finite
range for both of the coordinates conjugate to non-compact translation symmetries. We have
also discussed how to obtain the spin dependence of anomalous
dimensions of double trace operators and we found that it is given by a power
determined by the twist of the lowest twist operator that couples to each single trace
operator \correct .
  We also explained how the weak coupling
computation of the cusp anomalous dimension can be reduced to an
effective two dimensional QCD problem and argued that the cusp
anomalous dimension for  arbitrary representations   displays
Casimir scaling up to three loops. The arguments are made on the field theory side
and are valid for any conformal gauge theory regardless of the value of the coupling.
We have discussed also the extension to the non-conformal case for the case of the
Sudakov factor.

We then considered operators with high spin and charge in ${\cal N}=4$ super Yang Mills
and argued that in the limit where $J/\log S $ is finite we get
$\Delta - S = [ f + \epsilon( J/\log S) ] \log S $ \FrolovQE . We showed that when
$J/\log S$ is suitably small the computation of the function $\epsilon(J/\log S)$ reduces
to a computation in the bosonic O(6) sigma model. This relation gives us a way
of obtaining exact results for the worldsheet string theory. These can, hopefully, be
used to test the BHL/BES \refs{\bhl,\bes}
 prediction for the phase of the S-matrix at higher loops.

{\bf Acknowledgments}

  We would like to thank
 S. Ellis, G. Korchemsky, A. Manohar, G. Sterman,  E. Sokatchev and E. Witten
 for discussions.

This work   was  supported in part by U.S.~Department of Energy
grant \#DE-FG02-90ER40542. The work of L.F.A was supported by VENI
grant 680-47-113.

{\bf Note Added:} 

 After this paper appeared, \ref\BassoWD{
  B.~Basso, G.~P.~Korchemsky and J.~Kotanski,
  arXiv:0708.3933 [hep-th].
}    computed the strong coupling
expansion of the cusp anomalous dimension from the BES equation \bes , finding
exponentially small corrections that precisely agree with $m^2$, with $m$
given by eqn. \massexc . Since the cusp anomalous dimension has the
interpretation of an energy density, we expect this kind of corrections
besides the perturbative series in $1/\sqrt{\lambda}$.

\newsec{Appendix A: One loop computation for the cusp anomalous dimension}

In this appendix we outline the one loop computation of the cusp anomalous dimension using
the coordinates we introduced above. Of course, this is a well known computation that has
been done in many ways.
We perform the computation by considering a Wilson loop in the coordinates \newcoord .
We consider first a $U(1)$ theory, and start with a
configuration with an electric flux which is non-zero only  in the
$u,\chi$ directions, $F_{u\chi}$, which is constant due to the Bianchi identity.
We can write the action as
\eqn\action{S_{act}=-{1 \over 4 g^2}\int d^4x \sqrt{-g}F_{\mu
\nu}F^{\mu \nu}= {|F_{u\chi}|^2 \over 2 g^2} \int {d\psi du d\chi
d\sigma \over \cosh 2 \sigma} ={\pi^2 \over 2g^2}\int du d\chi
|F_{u\chi}|^2 }
where we have integrated over the circle parametrized by $\psi$ in \coordsta .
  The total energy density is
\eqn\En{{ f \over 2 } =  {\pi^2 |F_{u\chi}|^2\over 2g^2} }
 The quantization condition for $F$ can be obtained in the standard way after we say that
 fundamental charges couple as $e^{ i \int A}$ and compute the amount of flux that this
 charge generates. This gives $F_{u\chi} = {g^2\over \pi^2} $.
 We then obtain
 \eqn\finand{
 { f \over 2} = { g^2 \over 2 \pi^2 }
 }
 We can now consider the generalization to a $U(N)$ theory. In that case we get a similar
 result except that we get an additional factor of $N$ in \finand\ and there is an
 additional factor of two that comes from the conventional definition of the coupling.
 Alternatively, we can reduce the four dimensional theory to two dimensions along $\psi$
 and $\sigma$. This leads to a two dimensional theory with a coupling $g_2^2 = g_4^2/\pi^2$.
 Then we can use the two dimensional QCD result  \valef\ with $C_2 = N/2$
 for the fundamental representation so that we get the one loop value
 $f/2 = { g^2 N \over 4 \pi^2 }$.

 \subsec{Spin generators in the new coordinates}

As we see in the metric \newparm\ the
 generator that measures spin, $ -i \partial_\varphi$
  corresponds to an isometry in $AdS_3$.   We can now compute the
  form of this Killing vector in the other  $AdS_3$  coordinates
 \newcoord
 \eqn\gener{\eqalign{\Delta= {1 \over
2}\left(1+{\cosh 2\chi \over \cosh 2\sigma} \right) i \partial_u+{1
\over 2}\cosh 2\chi \tanh 2 \sigma i \partial_\chi + \cosh \chi \sinh
\chi
i \partial_\sigma \cr
S=  {1 \over 2}\left(-1+{\cosh 2\chi
\over \cosh 2\sigma} \right)i \partial_u+{1 \over 2}\cosh 2\chi \tanh
2 \sigma i \partial_\chi + \cosh \chi \sinh \chi
i \partial_\sigma}}
where we also indicated the form for $\Delta$.
We see   that $\Delta-S=i\partial_u$.
We see from these expressions that if we have a configuration which goes up to
some distance $\chi_0$, then its spin   scales as $S \sim e^{2 |\chi_0|}$.
We can understand this better if we compute the contribution to the spin of the
constant flux configuration discussed above.
Given
a general Killing vector of the form $V=i \xi^\mu \partial_\mu$ the
conserved current associated to it is given by contracting $\xi^\mu$ with the
stress tensor. Using the expression for the stress tensor for a gauge field we find,
after integrating over $\psi$ in \newcoord ,
\eqn\charges{\eqalign{\Delta={\pi \over g^2} \int d\sigma d \chi
{F_{u \chi}^2 \over \cosh 2\sigma}\left({1 \over 2}+{\cosh 2\chi
\over 2 \cosh 2\sigma} \right)\cr S={\pi \over g^2} \int d\sigma d
\chi {F_{u \chi}^2 \over  \cosh 2\sigma}\left(-{1 \over 2}+{\cosh
2\chi \over 2 \cosh 2\sigma} \right)}}
Of course, this flux configuration is not a good description of the quark-antiquark configuration
near $\chi \sim \chi_0$. However, we argue that the dynamical particles that we put at $
\chi \sim \chi_0$   still have a spin of order $S \sim e^{ 2 |\chi_0|}$.
In this fashion we connect the range of $\chi$ to the spin, via
$\Delta \chi =  \log S $.

\newsec{Appendix B:  Renormalization group and  evolution on the cylinder }

A conformal field theory on the plane is equivalent to a conformal field theory
on the cylinder and the spectrum of anomalous dimensions of operators on the plane corresponds
to the energy spectrum for the theory on the cylinder.
Now, suppose that we have a non-conformal field theory on the plane. For simplicity, imagine we
have a theory with a single coupling that runs $g^2(\mu)$. Then we would like to understand
what type of theory we get on the cylinder.

For simplicity, consider the Euclidean theory. Then the plane and the cylinder are related by
the following Weyl transformation
\eqn\metrplcyl{
ds^2_{R^4} = r^2 \left( {d\tau^2 } + d\Omega_3^2 \right) = r^2 ds^2_{R \times S^3 } ~,~~~~~~~
r= e^\tau
}
Thus the two metrics are related by $ds^2 = \Omega^2 d{s'}^2 $, $\Omega =r = e^\tau$.
 Let us imagine regularizing the
field theory on the plane with a cutoff $\Lambda$ so that the value
of the coupling at the cutoff
is constant. On the cylinder this leads to a field theory where the cutoff
$\Lambda' = \Omega(x) \Lambda$ depends on position.
 In our case, this leads to
a cutoff $\Lambda' = e^{\tau} \Lambda$  which depends on the Euclidean time direction along the
cylinder. The coupling constant on the cylinder is a constant at scale $\Lambda'$.
This can be related to the more conventional way of defining the theory on
the cylinder which uses a   fixed (time independent)
 cutoff $ \Lambda''$.
We can obtain the value of the coupling at $ \Lambda''$ by using the renormalization group
equation to evolve the coupling from the scale $\Lambda'(\tau)$ to the scale $\Lambda''$.
We can apply the ordinary flat space renormalization group equation as long as the coupling
is varying slowly at the scale of the cutoff. This condition reads ${ \partial_\tau \Lambda'/\Lambda'}
 \ll \Lambda'$. In our case this requires that $e^{\tau } \Lambda \gg 1$. This says
 that the cutoff $\Lambda'$ should be bigger than the inverse radius of the $S^3$.
   Thus, if we want
 to explore the theory at $\tau \to -\infty$ (or $r\to 0$), we need to take a smaller and smaller
 cutoff $\Lambda$, which is of course what we expect.

In conclusion,  we have a ``time'' dependent theory on the cylinder, where the
coupling constant depends on $\tau$. The time dependence of the coupling constant can be
computed exactly if we know the exact $\beta$ function.
Since we have a time dependent theory on the cylinder we have a time dependent Hamiltonian.
We can nevertheless diagonalize this Hamiltonian at each time and this   leads to the
scale dependent anomalous dimensions we have in a non-conformal theory $\Delta(g^2(\tau) )$.

For theories that have a gravity dual, it is useful to understand this also from the
gravity perspective.
Let us start with a five dimensional metric and scalar field
\eqn\ffsol{
ds^2  = w(z)^2 { dz^2 + dx^2 \over z^2 } ~,~~~~~~~~\phi(z)
}
We can now use the usual change of coordinate that takes us between the plane and the cylinder
in the $AdS$ case,
which is
\eqn\changc{
 { 1 \over z} = { e^{-\tau} \cosh \rho } ~,~~~~~~~~ { r \over z } = \sinh \rho
 }
 so that we end up with the metric and scalar fields
 \eqn\mmnew{
 ds^2 = w\left( { e^{\tau} \over \cosh \rho } \right)^2 \left[
 \cosh^2 \rho d\tau^2 + d\rho^2 + \sinh^2 \rho  d\Omega_3^2 \right]  ~,~~~~~~~~
 \phi\left({ e^\tau \over \cosh \rho }\right)
}
We now see that in the original metric \ffsol\ it is natural to impose a cutoff at
$z = 1/\Lambda$. This cutoff would then correspond to
$   e^{-\tau} \cosh \rho_c(\tau) = \Lambda$, where $\rho_c(\tau)$ is determined by this
equation. From the point of view of the new theory, we would say
that the cutoff is at $   \Lambda'\sim e^{\rho_c(\tau)}/2 \sim \Lambda e^\tau$
if $\rho_c$ is large. Thus we get
$\Lambda' = e^\tau \Lambda$ as we had in the general discussion. Notice that the
condition that the time variation was slow, translates into
the condition that $\rho_c \gg 1 $.
If we fix the cutoff at a time independent value
$\Lambda'' = e^{\rho''_c}/2$ we
get that the scalar field has a time dependent value given by \mmnew . In fact, we
get $\phi_c \sim \phi( 2 e^{\tau - \rho_c''})$.

\newsec{Appendix C:  O(N) sigma model }

In this appendix we recall some results for the $O(N)$ non linear
sigma model.
We consider the $O(N)$ sigma model in the presence of a chemical
potential $h$ coupled to one of the conserved charges (an $SO(2) \subset SO(6)$)
and we
 compute the free energy $f(h)=min_j[\epsilon(j)-j
h]$.
By dimensional analysis $f = h^2 { \cal F}(h/m)$, where $m$ is the mass gap.

Given the two particle $S-$matrix for the $O(N)$ $\sigma-$model
\eqn\Smatrix{\eqalign{
S(\theta)= & -{\Gamma(1+x)\Gamma(1/2-x)\Gamma(1/2+\Delta+x)\Gamma(\Delta-x)
\over \Gamma(1-x)\Gamma(1/2+x)\Gamma(1/2+\Delta-x)\Gamma(\Delta+x)}
\cr
 &x =  {i \theta \over 2\pi}~~~~\Delta={1 \over N-2}
 }}
The thermodynamic Bethe ansatz leads to an integral equation for the
free energy \HasenfratzAB . More precisely \eqn\freeen{f(h)=-{m \over 2\pi}
\int_{-B}^B \cosh(\theta)\rho(\theta) d\theta } where $\rho(\theta)$
satisfies the following integral equation with the following
boundary condition \eqn\inteq{\eqalign{\rho(\theta)-\int_{-B}^B
K(\theta-\theta')\rho(\theta')d\theta'=h-m \cosh(\theta)},~~~~~~~~~
\rho(\pm B)=0 } where $K(\theta)={1\over 2 \pi i} {d \over
d\theta}\log S(\theta) $.

In order to make a comparison with computations from the string
sigma model, one needs to consider the regime $h/m \gg 1$ which
corresponds to the weakly coupled region of the theory. This
 was considered in \HasenfratzAB  . Their analysis leads to the
following result for the free energy
\eqn\freelarge{\eqalign{
f(h)= &  -{h^2 \over 2}\left(-\beta_1 \log(h/m)+{\beta_2
\over \beta_1} \log\log(h/m)+c +\tilde{c}  {\log\log(h/m) \over
\log(h/m)}+... \right)
\cr
\beta_1= &-  {N-2 \over 2\pi}~,~~~~~~\beta_2= - {N-2 \over
4\pi^2}~,~~~~~~~~c ={N-2 \over 2\pi}\log{\left[\left({8 \over
e}\right)^{1 \over N-2}{e^{-1/2}\over \Gamma(1+{1\over
N-2})}\right]}
}}
where the notation is chosen to highlight the fact that
$\beta_{1,2}$ turn out to be
the one and two loop beta functions for the
$O(N)$ sigma model.
Next, we perform a Legendre transform and express $\epsilon=f(h)+j
h = j^2 { \cal E}(j/m)$ in terms of the charge density $j \equiv - f'(h)$.
 Starting from \freelarge\ it is
straightforward to iteratively solve for ${\cal E}(j/m)$
\eqn\Fexp{\eqalign{ {\cal E}(j/m)& ={1 \over -2 \beta_1 \log
j/m}+{1 \over - 2 \beta_1^3 \log^2
j/m}\left((\beta_1^2+\beta_2)\log\log j/m + \beta_1 c+\beta_1^2
\log{(-\beta_1)} \right)\cr & + {1 \over - 8 \beta_1^5 \log^3
j/m}\left[ 4(\beta_1^2 + \beta_2)^2 (\log\log j/m)^2- k' \log\log
j/m+const \right]+ \cr & + {\cal O}({1 \over \log^4 j/m})\cr k' &
\equiv 4\beta_1\left(\beta_1^3+2\beta_1\beta_2-2\beta_2 c
-\beta_1^2(2c+\tilde{c})-2\beta_1(\beta_1^2+\beta_2)\log{(-\beta_1)}\right)
}} The constant piece in front of ${1 \over \log^3 j/m}$ depends
on a higher order term, of the form $1/\log(h/m) $,
 in \freelarge\ which
  we have not computed.
In order to make a comparison with the results of \FrolovQE\ we
need to express our expansion parameters $j$ and $m$ in terms of
$y$   and the coupling constant $\lambda$ using \defx , \massexc ,
or \eqn\massexp{ m = k \lambda^{ - { \beta_2 \over 2 \beta_1^2} }
e^{ { \sqrt{ \lambda} \over 2 \pi \beta_1 } } } Expanding
$\epsilon(\lambda,x)$ as a power series on $\lambda$ we obtain
\eqn\lambdaexp{\eqalign{
\epsilon(\lambda,x)=&y^2 \left[
{\sqrt{\lambda} \over 4 \pi  }  +   { \beta_1 \over 2} \log(y)+k_1
 \right. +\cr & \left. +{\pi \over 4  \sqrt{\lambda} \beta_1^2  }\left(4
\beta_1^4 \log^2(y/(2 \pi k))+k_2 \log y+ (2 \beta_2^2+2\beta_1^3
\tilde{c} )\log(\lambda)+const \right)+...\right]
}}
 Where the
terms $k_1$ and $k_2$ entering at one and two loops are
\eqn\kone{\eqalign{2 k_1 = &
 - c - \beta_1 \log k + { \beta_2 \over \beta_1} \log(-2 \pi \beta_1)\cr k_2=&
4\beta_1^2 \left[\beta_1^2+\beta_2-2\beta_1 c  +2 \beta_1^2\log(2
\pi)+2\beta_2\log(-2\pi \beta_1) \right]}}

Now let us discuss some properties of the function ${ \cal F}$ in
\freelarge . This function has a structure of the form
\eqn\strucf{ {\cal F}(t) = f/h^2 = -{1 \over 2} \left( -\beta_1 t
+ { \beta_2 \over \beta_1} \log t + c + \sum_{n=1}^\infty
\sum_{m=0}^n a_{nm} { ( \log t)^m \over t^n } \right) } This
structure is dictated by the structure of perturbation theory.
Moreover, all the coefficients for terms involving logarithms,
$a_{nm}$ with $m>0$,
 are determined in terms of lower order coefficients,
$a_{n'm}$ with $n'<n$. In particular,
 the constant $\tilde c$ is determined by the structure
of perturbation theory
\eqn\ct{\tilde{c} =-{\beta_2^2 \over
\beta_1^3 }}
Notice that with this value of $\tilde c$ the term involving a $\log \lambda$ in
\lambdaexp\ disappears. In general, the logarithmic terms in \strucf\ are fixed by
demanding that we can express the answer in terms of a  power series expansion in
terms of the effective coupling $ \bar{g}^2(\mu)$ of the sigma model, with the additional
condition that the
  dependence of the coupling constant $ \bar{g}(\mu)$ on the scale
$\mu$ is described by the Callan-Symanzik equation \eqn\cz{ \mu
\partial_\mu  {\bar g}^2(\mu)=\beta_1 \bar{g}^4(\mu) + \beta_2
\bar{g}^6(\mu)+... = \beta(\bar g^2(\mu) ) } where $\beta(g^2)$
also has a power series expansion in $\bar{g}^2$. This beta
function equation can be solved as
\eqn\betfneq{
 { 1 \over {\bar
g}^2} = - \beta_1 \log(\mu/\Lambda) + { \beta_2 \over \beta_1 }
\log(\log{\mu/\Lambda} ) + \cdots
 }
  where $\Lambda$ is the
dynamical scale of the theory. If we  take the renormalization
scale at $\mu = h$, then we can think of $\log(h/\Lambda) = t
+$constant. In that case we can solve the equation as \eqn\formt{
t + const = { 1 \over - \beta_1 \bar{g}^2} + { \beta_2 \over
\beta_1^2 } \log (1/\bar{g}^2) + o(\bar{g}^2 ) } We see that as we
take ${\bar g}^2$ on a small circle around the origin $ \bar{g}^2
\to
 \bar{g}^2 e^{ 2 \pi  i }$,
then $t$ performs a circle around infinity, but in addition we get a shift,
$t \to t e^{- 2 \pi i } - { \beta_2 \over \beta_1^2} 2 \pi i $.
When we say that ${\cal F}$ has a power series expansion in $\bar{g}^2$ we are saying that each
term is invariant under this shift. However, $t$ is not invariant and this one way to see
that we need the logarithmic terms in \strucf , with   coefficients determined by the
lower order terms. All the logarithmic terms in \strucf\ would vanish if $\beta_2$ were zero.

Now let us turn to the question of determining the leading logarithmic terms at each loop
order in the $1/\sqrt{\lambda}$ expansion. One can determine such terms by performing
manipulations similar to the ones performed above. However, it is also nice to see more
directly how they are determined by using the renormalization group equations.
For that purpose we imagine that we compute the tree level and one loop expressions for
${\cal F}$ using perturbation theory. The simplest answer is obtained by choosing $\mu = h$
in which case we get
\eqn\simplf{
f = h^2 {\cal F} ~,~~~~~
{\cal F}  = - { 1 \over 2 } \left[ { 1 \over \bar{g}^2(h) } + a  + o(\bar{g}^2) \right]
}
where $a$ is a constant that we  will  fix momentarily.
We now run the coupling from the scale $h$ to the scale 1 corresponding to the UV cutoff
where the coupling is $\bar g_0^2 = \bar g^2(1)=
 { 2 \pi \over \sqrt{\lambda} } ( 1 + { \hat c \over \sqrt{\lambda}}) $. The
 coupling $\bar g^2{1}$
 is the coupling run to the scale 1 in the $O(6)$ theory and
 constant $\hat c$ is an unknown threshold correction. We can use the solution
\formt . We then see that we can express the coupling at scale $h$ as
\eqn\newcoup{
{ 1 \over {\bar g}^2(h)} = { 1 \over \bar{g}_0^2} \left[ 1 + \bar
z + \bar{g}_0^2 {\beta_2 \over \beta_1} \log(1 + \bar z)
\right] ~,~~~~~\bar z \equiv - \beta_1 \bar{g}_0^2 \log h
}
where we neglect higher orders in $\bar g_0^2$ but kept all orders in $\bar z$.
We now compute $j$ to find
\eqn\expforj{
j = - { \partial f \over \partial h } = h { 1 \over \bar{g}_0^2} \left[
1 + \bar z + \bar{g}_0^2 {\beta_2 \over \beta_1} \log(1 +\bar z) + \bar{g}_0^2 a -  \bar{g}_0^2 {
\beta_1 \over 2}
\right]
}
We then solve for $h$ as a function of $\bar y \equiv  \bar{g}_0^2 j$. This gives
\eqn\energ{
h = { \bar y \over 1 + \hat  z + \bar{g}_0^2 \beta_1 \log(1 + \hat z) + \bar{g}_0^2
{\beta_2 \over \beta_1} \log(1+ \hat z)  +
a {\bar g}_0^2 - \bar{g}_0^2  { \beta_1 \over 2} } ~,~~~~\hat z \equiv - \beta_1
\bar{g}_0^2 \log {\bar y}
}
The expression for the energy is then
\eqn\energyexp{
\epsilon(  y) ={ {\bar y}^2 \over 2 \bar{g}_0^2 } \left[
{ 1 \over (1 + \hat z) } - {  a \bar{g}_0^2   +
\bar{g}_0^2\beta_1 ( 1 + { \beta_2 \over \beta_1^2 } )
\log(1 + \hat z) \over (1 + \hat z)^2 }
\right]
}
We now recall that $y = {2 \pi \over \sqrt{\lambda} } j$. Then
 $ \bar y = y (1 + { \hat c \over \sqrt{
\lambda} })$. Similarly we can  define
$z = - \beta_1 { 2 \pi \over  \sqrt{\lambda}} \log y $, so that
$ \hat z = z( 1 + { \hat c \over \sqrt{\lambda} } )$. We can then write
\energyexp\ as
\eqn\newenexp{
\epsilon(  y) ={ {y}^2 \sqrt{\lambda} \over4 \pi  } \left[
{ 1 \over (1 +  z) } - { k_3 { 2 \pi \over \sqrt{\lambda}} +
{ 2 \pi \over \sqrt{\lambda} } \beta_1 ( 1 + { \beta_2 \over \beta_1^2 } )
\log(1 +  z) \over (1 +  z)^2 }
\right]
}
where $k_3$ is a combination of the unknown parameters. $k_3$ can be  fixed by
  performing the one loop expansion of \newenexp\ and
matching to the results in \FrolovQE . The one loop expansion of \newenexp\ is
\eqn\expanso{
\epsilon_{0+1} = y^2 \left[ { \sqrt{\lambda} \over 4 \pi}  + { \beta_1 \over 2 } \log y -
{ k_3 \over 2 } \right]
}
Comparing this to \xexpr\  we
find that $ -k_3/2 = 3/(4\pi)$. Inserting this in \energyexp\ and
using  the values of $\beta_1, \beta_2$ for the O(6) model \freelarge , we find
\eqn\expansnew{
\epsilon = { y^2 \sqrt{\lambda } \over 4 \pi }
\left[ { 1 \over (1 + z) } + { 3 \over \sqrt{\lambda} }{
1 + \log(1+z) \over (1 + z)^2 }
\right] ~,~~~~~~z = { 4 \over \sqrt{\lambda} } \log y
}
The first term gives the leading log terms and the second gives the subleading ones that
we had quoted in \twoleading .

Let us now make some final remarks. Obviously,
  in the perturbative region we can express
 the free
energy in terms of the running coupling constant $\bar{g}(\mu)$, at
some scale $\mu$
\eqn\fg{f(h)=h^2 \left({{\cal K}_1(h/\mu) \over
\bar{g}^2(\mu)}+{\cal K}_2(h/\mu) +{\cal K}_3(h/\mu)
\bar{g}^2(\mu)+...\right)}
Then the functions $K_i$ are determined by the RG equations up to a constant.

The
relation between the mass gap $m$ and the scale defined in the $ \overline{MS}$
scheme via the equation \betfneq\ , $\Lambda_{\overline{MS}}$,
 was computed in
\HasenfratzAB\
\eqn\mvsl{m=\left({8 \over e} \right)^{1 \over N-2}{1 \over
\Gamma(1+{1 \over N-2})}\Lambda_{\overline{MS}}}

\listrefs

\bye